\documentstyle[sprocl,epsfig,11pt]{article}
\def\be{\begin{equation}}
\def\ee{\end{equation}}
\def\bea{\begin{eqnarray}}
\def\eea{\end{eqnarray}}
\begin{document}
\title{THE AMANDA EXPERIMENT - status and prospects\\
for indirect Dark Matter detection\footnote{
Talk given at {\em Identification of Dark Matter}, Sheffield, September 1996,
to appear in the Proceedings} }
\author{{\bf by L. Bergstr\"om}\footnote{E-mail: lbe@physto.se}}
\author{for the AMANDA Collaboration: }
\address{}

\author{ D. M. Lowder,  T. Miller\footnote{At Bartol Research Institute,
University of Delaware, Newark,Delaware }, D. Nygren\footnote{ At Lawrence 
Berkeley
National Laboratory, Berkeley, CA 94720, USA }, P. B. Price, and A. Richards }
\address{ University of California at Berkeley, Berkeley,
CA 9472,USA}
\author{ S. W.  Barwick, P. Mock, R. Porrata, E. Schneider and G. Yodh}
\address{ University of California at Irvine, Irvine, CA
92717,USA }
\author{ E. C. Andr\'es, P. Askebjer, L. Bergstr\"om, 
A. Bouchta, E. Dahlberg, P.Ekstr\"om, B.~Erlandsson, A.~Goobar,
 P. O. Hulth, Q. Sun, and C. Walck }
\address{ Stockholm University, Box 6730 S-113 85
Stockholm, Sweden }
\author{ S. Carius, A. Hallgren, and H. Rubinstein }
\address{ Uppsala University, Box 535,S-75121
Uppsala,Sweden }
\author{ K. Engel, L. Gray, F. Halzen, J. Jacobsen, V. Kandhadai, I. Liubarsky,
R. Morse, and S. Tilav }
\address{ University of Wisconsin, Madison, WI 53706,USA }

\author{H. Heukenkamp, S. Hundertmark, A. Karle, Th. Mikolajski, 
T. Thon, C. Spiering, O. Streicher, Ch. Wiebusch and
 R.Wischnewski }
\address{ DESY-IfH Zeuthen, D-15735 Zeuthen, Germany }

\maketitle\abstracts{ At the AMANDA South Pole site, four new holes were
drilled to depths 2050 m to 2180 m and instrumented with 86 photomultipliers 
(PMTs) at depths 1520-2000 m. 
Of these PMTs 79 are working, with 4-ns timing resolution and noise
rates 300 to 600 Hz. Various diagnostic devices were deployed and are working.
An observed factor 60 increase in scattering length and a sharpening of the
distribution of arrival times of laser pulses relative to measurements at
800-1000 m showed that bubbles are absent below 1500 m. 
Absorption
lengths are 100 to 150 m at wavelengths in the blue and UV to 337 nm. Muon
coincidences are seen between the SPASE air shower array and the AMANDA PMTs at
800-1000 m and 1500-1900 m. The muon track rate is 30 Hz for 8-fold triggers
and 10 Hz for 10-fold triggers. The present array is the nucleus for a future
expanded array.
The potential of AMANDA for SUSY dark matter search through the detection
of high-energy neutrinos from the centre of the Sun or Earth is discussed. }

\section{Introduction } The deep Antarctic ice is the purest, most transparent
of all natural solids. As a site for a high-energy neutrino observatory it has
a number of advantages compared to deep sea water. It consists of compressed
pure-H$_2$O
snow with the lowest contamination by aerosols and volcanic dust of any place
on Earth, and it contains neither bioluminescent organisms nor radioactive
${^{40 }K}$. Before the AMANDA collaboration began to measure the optical
properties of ice at the South Pole, one could have listed a number of
potential drawbacks: No one had ever drilled a hole deeper than 349 m at South
Pole; the depth at which air bubbles completely transform into solid crystals
of air hydrate clathrate was not known; the absorption length of light in ice
was thought to be shorter than in sea water~\cite{bd,be};
and the effects of dust, traces of marine salt, traces of natural acids, and
birefringence of polycrystalline ice on scattering of light in ice at South
Pole had not been studied. These issues have now been addressed.

\begin{figure}[h]
         \begin{center}
            \mbox{\psfig{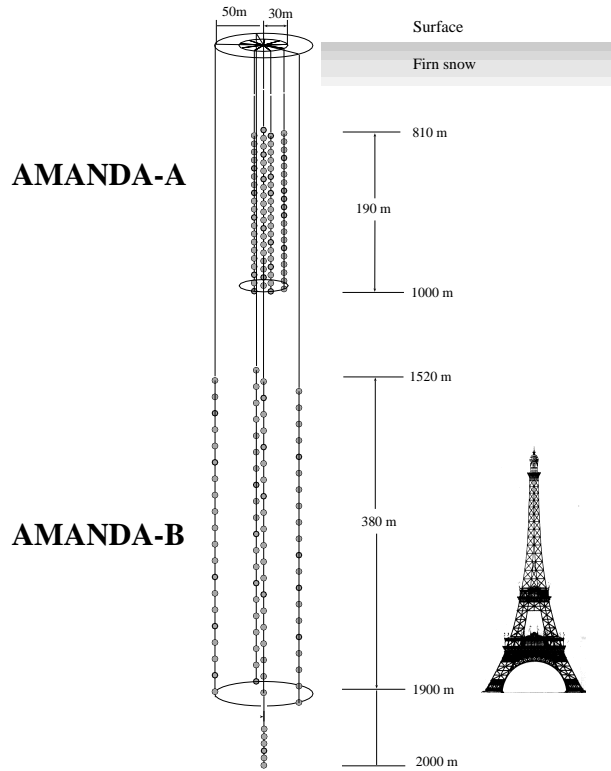}}
         \end{center}
         \caption{Schematic view of the AMANDA detector.}
         \label{amandaab}
\end{figure}

\section{Results from the AMANDA-A Array at 800 m to 1000 m }The successful
deployment of the four-string AMANDA-A array with photomultipliers (PMTs) 
was the
first step toward demonstrating that the South Pole ice is a suitable site for
a high-energy neutrino observatory~\cite{ro}. With a hot-water drill, four
holes 60 cm in diameter were created during the 1993-94 season and
instrumented with 80 PMTs spaced at 10-m intervals from 810-1000 m. To
measure
optical properties, a laser in a laboratory at the surface above
the holes was
used to send nanosecond pulses down any of 80 optical fibers to emitting balls
located near each PMT. From the distributions of arrival times at neighboring
PMTs, it was possible to determine separately the absorption length $\lambda_a$
and effective scattering length, $\lambda_e = \lambda_s /(1 - <\cos \theta>)$, at
wavelengths from 410 to 610 nm. Here $<\cos \theta >$ is the mean cosine of the
scattering angle. Because $\lambda_e  \ll \lambda_a$ the data fitted an
expression for three-dimensional diffusion with absorption. 
        In contrast to laboratory ice, for which $\lambda_a$  was
reported~\cite{bd,be} to be $<25$ m at all wavelengths and $\lambda_a$
= 8 m at the wavelength for maximum quantum efficiency of a PMT, we found
values of $\lambda_a$ exceeding 100 m at wavelengths less than ~480 nm and
values exceeding 200 m at wavelengths less than about 420 m~\cite{sa,sb}.
         We found that, independent of wavelength, $\lambda_e$ increased
monotonically from 40 cm at 820 m to 80 cm at 1000 m. We interpreted this
result as evidence for scattering by air bubbles with size much larger than the
wavelength of light, with the size and number density of bubbles decreasing
with depth.

\section{Technical Aspects of the AMANDA-B 1995-96 Drilling Season }

        New drilling equipment, operating at a power of 1.9 MW, used water
emerging at 75 C to drill at a rate up to 1 cm/s. It required a time of no
more than 4 days to melt a 60-cm-diameter cylinder of ice to a depth of 2000 m.
Due to a late start and several problems associated with commissioning the new
equipment, only four holes were drilled, of which one reached a depth of
2180~m. It took typically 8 hours to remove the drill and water-recycling pump
from
a completed hole. The rate of refreeze was 6 cm decrease in diameter per day,
which easily allowed time to mount PMTs and other devices on cables, to lower
the cables, and to route the upper ends of the cables into the AMANDA
building in time to monitor the entire refreezing process.
        The diagnostic devices included four inclinometers to measure shear vs
time, thermistors to measure temperature vs depth, and pressure gauges to
follow the refreezing. The measurements of temperature at three depths,
together with previous measurements, confirmed the validity of a model of
temperature vs depth. At the greatest depth, the temperature of the ice was 
-31 C, about 20 degrees warmer than at the surface. 
        Of the 80 PMTs on the coaxial cables 73 survived the deployment and
refreezing, and all of the 6 PMTs in the bottom 100 m of a prototype
twisted-pair cable are working. With a total of 152 operating PMTs, the overall
success rate is greater than 90$\%$ 
for phototube deployment on AMANDA-A and B. No
PMTs have failed since refreezing of the ice. The mean time to failure is
inferred to be $>$200 years per PMT. 
        With no local amplification, the analog signals are preserved, though
broadened, in transmission along a 2-km coaxial cable, with a standard
deviation of better than 4 ns in timing resolution. Of this, 2.5 ns is due to
the resolution of the optical fiber itself. 
        The noise rates of the 8-inch Hamamatsu PMTs are in the remarkably low
range of 350 Hz 
to 600 Hz. The twisted-pair cable has a significantly shorter rise
time than the coaxial cable and requires front-end amplifier gains of only 30
instead of 100. A great advantage of the twisted-pair cable is that a single
cable can supply 36 PMTs instead of only 20.
        At the surface, the new ADCs and new amplifiers are working as well as
hoped. A newly installed trigger logic to search for gamma-ray bursts and
supernovas at timescales of milliseconds and seconds is operating with 64
optical modules at 0.8 
km to 1 km depth and with 79 optical modules at 1.5 km to 2.0
km. Data are being taken by our Bartol colleagues with two radio receivers at
depths of 150 m and 280 m, their aim being an initial evaluation of a method
of detecting Cherenkov radiation at radio frequencies by ultrahigh energy
cascades in the ice.
        The YAG laser in the surface laboratory provides tunable pulses at 410
to 610 nm with only 10 dB loss down the optical fibers. A pulsed nitrogen
laser (337 nm) at a depth of 1820 m, held at a temperature of plus 24 C, is
operating flawlessly. Pulsed blue LED beacons with filters for 450 and 380 nm
emission are operating at various depths. DC lamps at 350 nm, 380 nm, and
broadband are also operating.

\section{Physics Results } 

\subsection{Ice properties at 1500 - 2000 m }

        The burning issue -- whether the bubbles are still present at depths
1500 m to 2000 m -- is now settled. Preliminary analysis show $\lambda_e$
in the range 25-30 m which is  two orders of magnitude
greater than at 800-1000 m. The value of
$\lambda_e$ shows no strong dependence on wavelength nor on 
depth. Because
$\lambda_e$ is comparable to the spacing between neighboring optical modules,
many of the photons from one emitter have undergone zero or few scatters before
reaching a PMT. Thus, the analytic expression for diffusion with absorption is
inapplicable (because the photons are not in the diffusion regime). Our present
approach is to use Monte Carlo modeling and statistical techniques to find the
best values of $\lambda_a$ and $\lambda_e$ for each combination of emitter,
receiver, and wavelength.
        The large absorption length of ice in the blue and UV wavelength
regimes is confirmed by the AMANDA-B data. At 337 nm, $\lambda_a$ is of order
100 m, which is astonishing in view of the fact that $\lambda_a$ is only a few
meters for lake water and ocean water, and was reported to be only 5 m for
laboratory ice. At wavelengths in the blue, values of $\lambda_a$ significantly
longer than 100 m are being inferred from the data.
        Comparison of the data on $\lambda_a$ at 1500 m to 2000 m with data at
wavelengths 410 nm to 610 nm and at depths 810 m to 1000 m suggests that the
concentration of absorbing dust is greater at the greater depths. This is
consistent with our observation~\cite{sa,sb} 
that, at short wavelengths where
$\lambda_a$  is most sensitive to dust, $\lambda_a$  is constant at depths
800-900 m and decreases at depths 900-1000 m. Our interpretation is that the
ice at 800-900 m was formed in the post-glacial Holocene period ($<$13,000 
years BP)
where the dust concentration has been remarkably low, and that the ice at
900-1000 m was formed near the end of the most recent ice age, with a peak in
the dust concentration 17,000 years BP.
        Despite the larger concentration of dust at 1500-2000 m, the absorption
lengths and scattering lengths are acceptably long, allowing us to move
forward with plans for an extension of the AMANDA array.

\subsection{Data taking }

    We are now continously taking data with AMANDA-A and AMANDA-B. The rate in
AMANDA-B is
30 Hz for 8 triggers (at least 8 PMTs) within 2 microseconds, and is 10 Hz
for 10 triggers. Coincidences of
tracks of energetic muons between the surface SPASE array, AMANDA-A, and
AMANDA-B is also registrated. 
We are presently studying pattern recognition and muon track
reconstruction, with the goal of finding upward-going neutrino candidates. The
long tails of the timing distributions for muon tracks can be greatly narrowed
by cutting on two photoelectrons (p.e.), 
leading to much longer effective scattering
lengths than the 25-30 m values measured for single p.e. signals. 

In addition to the muon triggers both AMANDA-A and AMANDA-B are sampling the
total PMT noise rate.
The low counting rate of all PMTs could e.g. be increased
by Cerenkov light from interactions of low energy (few MeV)
neutrinos from stellar collapses.
A stellar collapse, similar to SN1987A and at 8kpc
distance, would yield about 100 counts per PMT in AMANDA-B.
This is large enough to already regard AMANDA-B with its only 86 PMTs
as a detector for neutrinos from supernovae up to the center of our galaxy.

\section{Future plans and prospects for Indirect Dark Matter detection}

        By any measure the 1995-96 AMANDA expedition has been a great success.
The large values of $\lambda_a$  mean that experiments that require large
transparent volumes without the need for tracking are already very effective.
These include searches for neutrinos accompanying gamma-ray bursts and
accompanying supernova explosions. The values of  
$\lambda_e$ of 25 m to 30 m are acceptably long, provided the spacing of 
PMTs is optimized. 
 
\begin{figure}[h]
         \begin{center}
            \mbox{\psfig{file=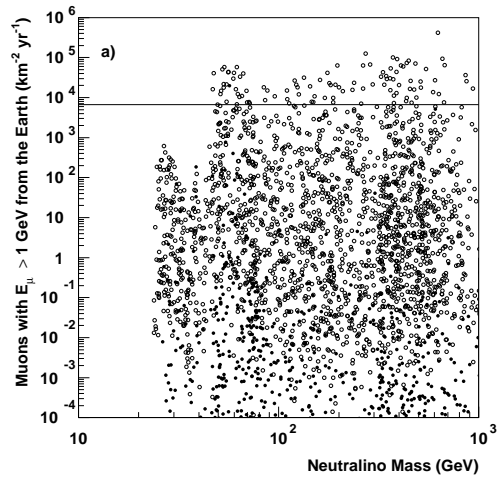,height=8cm,
                        bbllx=-80pt,bblly=0pt,bburx=558pt,bbury=470pt}}
         \end{center}
         \caption{The indirect detection rates from
  neutralino annihilations in the Earth  versus the
  neutralino mass. The
  horizontal line is the Baksan limit \protect\cite{baksan}. For 
  details, see \protect\cite{beg}.}
         \label{earth}
\end{figure}

            We are now preparing 6 new strings to be deployed during 96/97 in
the AMANDA-B detector. By changing to twisted pair cables for transmitting the
PMT signals to surface we are able to have 36 optical modules per string. 
We are also going to test an analog optical transmission of the PMT signals
over optical fibers allowing the "true" PMT signal at
surface. 

With this enlargement of the detector and with denser spacing
between the PMTs, we anticipate to get a pointing accuracy good enough
to start exploring interesting directional astrophysics. 
One of the most 
favourable 
signals to search for experimentally will be high-energy neutrinos from
the centre of the Earth. This process will have the lowest background from
atmospheric neutrinos, and also the least problem with the rejection of
downward-going muons from above, which for more horizontal directions could
mask an eventual signal. The effective area for vertical upward-going
muons should be of
the order of $10^4$ m$^2$ (the exact value will depend on various experimental
considerations, as will the threshold energy). In fact, at this time we
see no technical obstacle that would prevent a gradual increase of the 
effective area by another one or two orders of magnitude by just adding
more strings over, say, a 5-year period.

The detection of a high-energy neutrino signal from the centre of the 
Earth or the Sun 
would be a ``smoking gun'' for neutralinos or other WIMPs. The 
idea (see \cite{jkg} for a comprehensive review) is that WIMPs in the halo
occasionally scatter in the interior of the Earth
 or Sun and lose enough energy to be
gravitationally trapped. They then gradually fall to the centre where they
accumulate. Since neutralinos are their own antiparticles, they will 
annihilate when they encounter each other. The annihilation
 products
(fermions, gauge bosons and Higgs particles) will generate neutrinos of, 
typically, tens to hundreds of GeV energy (depending on the neutralino mass)
that can be detected by AMANDA. If no signal is found, useful limits on
the parameters of the SUSY model (see \cite{sandro}) can be put.

In Fig.\,2 (from \cite{beg}) is shown the prediction from a scan in SUSY
parameter space for the upward-going muon rate from neutralino annihilation
in the Earth. Also shown is the present upper limit from the Baksan
neutrino telescope \cite{baksan}, which has an area of only around 
100 $m^2$. As can be seen, by improving the Baksan bound by two orders of
magnitude one would probe a non-negligible fraction of SUSY parameter space.

{\section*{Acknowledgments}}

        The AMANDA collaboration is
 indebted to the Polar Ice Coring Office and to Bruce Koci for
the successful drilling operations, and to the National Science Foundation
(USA),  the Swedish National Research Council, the K. A. Wallenberg
Foundation and the Swedish Polar Research Secretariat. L.B. was sponsored
by the Swedish National Research Council and the European Union
Theoretical Astroparticle Network under contract No.~CHRX-CT93-0120 
(Direction G\'en\'erale 12 COMA).

\vskip .3cm
{

\end{document}